\documentclass[useAMS,usenatbib]{mn2e} 
\usepackage{graphicx}

\usepackage{times}

\defcitealias{LedouxC_03a}{LPS03}

\def\bsp_small{\vspace{0.5cm}\small\noindent This paper has been typeset
from a \TeX / \LaTeX\ file prepared by the author.}

\title[Cosmological evolution of heavy element and H$_2$
   abundances]{Cosmological evolution of heavy element and H{\boldmath$_2$}
   abundances} \author[S. J. Curran et al.]{S. J. Curran$^1$\thanks{E-mail:
   sjc@phys.unsw.edu.au}, J. K. Webb$^1$, M. T. Murphy$^2$ and R. F.
   Carswell$^2$\\
   $^1$School of Physics, University of New South Wales,
   Sydney NSW 2052, Australia\\
   $^2$Institute of Astronomy, , University of Cambridge, Madingley Road,
   Cambridge CB3 0HA, UK}

\begin{document}

\date{Accepted ---. Received ---; in original form ---}

\pagerange{\pageref{firstpage}--\pageref{lastpage}} \pubyear{2004}

\maketitle

\label{firstpage}

\begin{abstract}
Spectroscopic observations of distant quasars have resulted in the
detection of molecular hydrogen in intervening damped Lyman-$\alpha$
absorption clouds (DLAs).  We use observations compiled from different
experimental groups to show that the molecular hydrogen abundance exhibits
a dramatic increase over a cosmological time period corresponding to 13\%
to 24\% of the age of the universe.  We also tentatively show that the
heavy element abundances in the same gas clouds exhibit a faster and more
well-defined cosmological evolution compared to the general DLA population
over the same time baseline.  We argue that this latter point is
unsurprising, because the general DLA population arises in a wide variety
of galaxy types and environments, and thus a spans broad range of ISM
gas-phases and abundances at the same cosmic time.  DLAs exhibiting H$_2$
absorption may therefore circumvent this problem, efficiently identifying a
narrower class of objects, and provide a more sensitive probe of
cosmological chemical evolution.
\end{abstract}


\section{Introduction}\label{sec:intro}

The detection of cosmological chemical evolution is a cornerstone of modern
cosmology, providing empirical details about galaxy formation and
evolution, and independent supporting evidence of the big
bang. High-precision measurements of chemical abundances at early
cosmological epochs come from damped Lyman-$\alpha$ absorption systems
(DLAs), arising in distant galaxies which intersect the lines of sight to
high-redshift quasars. DLAs are easily detected in quasar spectra due to
their particularly high neutral hydrogen column densities [$\log_{10}
N($H{\sc \,i}$) \ge 20.3{\rm \,atoms\,cm}^{-2}$ by definition, though Voigt
profile damping wings are seen at lower $N($H{\sc \,i}$)$].  Generally,
they comprise a clumpy distribution of clouds, spread typically over a few
hundred ${\rm km\,s}^{-1}$.  Detection levels for heavy element transitions
in DLAs are $\sim$\,7 orders of magnitude below that for $N($H{\sc \,i}$)$.
High resolution optical spectra reveal transitions from many redshifted UV
species, allowing detailed heavy element abundance patterns to be studied.
The metallicity of DLAs -- defined as the heavy element abundance with
respect to hydrogen, relative to that of the solar neighbourhood: ${\rm
[M/H]} \equiv \log_{10}[N({\rm M})/N({\rm H})] - \log_{10}[N({\rm
M})/N({\rm H})]_\odot$ -- ranges over 2 orders of magnitude and is usually
$\sim$\,0.1\,solar or lower.  The presence of dust in DLAs is revealed by
the relative depletion of refractory elements (e.g.~Fe) with respect to
those thought not to readily deplete onto dust grains
\citep[e.g.~Zn;][]{PettiniM_97a}, and by the suppression of blue light from
the background quasar \citep[e.g.][]{FallS_89a}.  These spectroscopic observations
demonstrate that DLAs arise along lines of sight through distant galaxies
but they do not disclose details such as the galaxy's morphology,
luminosity, mass or age.

Evidence for an increase in DLA metallicity with cosmic time has emerged
only gradually over the past few years from a number of different studies
\citep{PettiniM_95a,LuL_96b,VladiloG_00a,KulkarniV_02a,ProchaskaJ_03b}. The
latter reference provides the most compelling statistical evidence so far
since, using the full sample of 125 DLAs, one sees a strong
anti-correlation between [M/H] and absorption redshift, $z_{\rm abs}$:
Kendall's $\tau$ is $-0.32$ with an associated probability of $P(\tau)\la
10^{-7}$. The need for such a large sample is due to the $\sim$\,2\,dex
scatter in [M/H] at a given epoch.

The types of galaxies responsible for DLAs can be investigated using direct
imaging methods or comparison of spectroscopic data with models. The
kinematics of optical resonance absorption transitions in low redshift DLAs
appear consistent with 2-component models: a rotating disk plus an extended
halo \citep{BriggsF_85a}.  A recent detailed radio study of one particular
DLA at $z_{\rm abs} = 0.437$ clearly shows that at least {\it some} DLAs
arise in the rotating disks of ordinary spiral galaxies
\citep{BriggsF_01a}.  High redshift ($z_{\rm abs} \sim 3$) DLAs have been
interpreted similarly, using detailed kinematic modelling to argue that
they are rotating ensembles of clouds with further evidence for this
interpretation coming from comparison of abundance ratios in DLAs and the
thick component of our Galaxy's disk \citep{ProchaskaJ_97b,WolfeA_01a}.

On the other hand, direct imaging of DLAs at $z_{\rm abs} < 1.5$ reveals
the hosts to be a mix of irregulars, spirals and low surface-brightness
galaxies \citep[LSBs; e.g.][]{LeBrunV_97a,RoaS_03a}, a result further borne
out by a blind 21-cm emission survey at $z=0$ \citep{Ryan-WeberE_03a}. The
observed number of DLAs per unit redshift interval and the $N($H{\sc
\,i}$)$ distribution suggest that DLAs at $z_{\rm abs} < 2$ are a roughly
equal mix of LSBs and spirals; at higher redshift they are more likely to
be dwarfs (that subsequently merge) since there are too few precursors of
present-day disk galaxies to explain the data
\citep{BoissierS_03b,BoissierS_03a}. Fitting simple chemical evolution
models to an observed $N($H{\sc \,i}$)$--[Zn/H] distribution suggests that
at redshifts $z_{\rm abs} \sim 2.5$, about half of all DLAs are due to
dwarf galaxies, while at $z_{\rm abs} < 1$, giant galaxies may dominate the
population \citep{BakerA_00a}. H{\sc \,i} 21-cm absorption measurements
also hint that dwarfs could dominate at high-$z$ \citep{KanekarN_03a}.


Further work is clearly needed to generate an accurate inventory of DLA
host galaxy types as a function of cosmic epoch.  Nevertheless, the
evidence for a broad mix of galaxy types is already compelling, if not
conclusive.

\section{Our Conjecture}

It is well established from studies of local galaxies
that chemical abundances depend on morphology, luminosity and location
within a galaxy. The diversity and evolution of the DLA host population
will further complicate and distort estimates of chemical evolution over
cosmological timescales. The DLA data themselves thus suggest that we may
derive greater insight by attempting to disentangle these different galaxy
types of DLA absorbers, rather than assuming all DLAs can be classified
into one sample.  A more sensitive test, or at least complementary test,
for chemical evolution may be arrived at if we can select a subset of DLAs
which arise in a {\it narrow range of physical conditions}.

One possibility, which we suggest here, is to target those DLAs which
exhibit molecular hydrogen absorption.  There are currently 8 examples of
such H$_2$-bearing DLAs at redshifts $z_{\rm abs} > 1.9$. None of the
absorbing galaxies have been identified directly through imaging, though
hints at the galaxy types come from a comparison with our own Galaxy and
its satellites.  The mean molecular hydrogen fraction, $f \equiv 2N({\rm
H}_2)/[2N({\rm H}_2)+N($H{\sc \,i}$)]$, seen along lines of sight through
our the Galaxy is $\log_{10}f \sim -1$, and for the SMC and LMC it is
$\log_{10}f\sim -2$. DLA detections lie in the range from $-6 < \log_{10}f
< -2$ (upper limits on non-detections are $-7 < \log_{10}f < -5$,
\citealt*{LedouxC_03a}, hereafter LPS03) but 2/3 of the detections are in
the range $-4 < \log_{10}f < -2$.  As discussed in
\citetalias{LedouxC_03a}, $>$\,90\% of the sightlines through our Galaxy
(and through the SMC) intercept H$_2$ clouds.  For the LMC, only 50\% of
sight-lines contain H$_2$ at the same strength.  DLAs exhibit H$_2$
absorption $\sim$\,20\% of the time \citepalias{LedouxC_03a}, and in this
respect are more similar to the LMC.  It is therefore possible that
H$_2$-bearing DLAs may arise predominantly in dwarf galaxies rather than
massive spirals. However, given the rapid apparent molecular fraction
evolution in Fig.\,1, this interpretation is speculative.

\section{Current observational evidence}

\begin{table*}\label{tab1}
\centering
\vspace{-1mm}
\begin{minipage}{\textwidth}
\caption{Summary of QSO and absorption line measurements for the 8
H$_2$-bearing DLAs considered here. Measurements of neutral hydrogen column
density [N(H{\sc \,i})], H$_2$ column density [N(H$_2$)], relative iron
abundance [Fe/H], and metallicity [M/H] are from \citet{LedouxC_03a}
(LPS03) unless otherwise noted. The QSO magnitudes given are approximate
$V\!$-band values. `S/N (H$_2$)' refers to approximate signal-to-noise
ratio per $\sim\!2{\rm \,km\,s}^{-1}$ pixel for the continuum in the
vicinity of the most clearly detected H$_2$ absorption lines. `M' refers to
the element used to calculate the metallicity.}
\begin{tabular}{lcccccccccc}\hline
QSO        & $z_{\rm em}$ & Mag.     & $z_{\rm abs}$ & $\log_{10}\!\left[\frac{{\rm N}({\rm H\,I})}{{\rm cm}^{-2}}\right]$ & $\log_{10}\!\left[\frac{{\rm N}({\rm H}_2)}{{\rm cm}^{-2}}\right]$ & $\log_{10}f$            & S/N (H$_2$)     & [Fe/H]             & [M/H]              & M \\\hline
0000$-$263 & $4.10$       & $18$     & $3.390$       & $21.41 \pm 0.08^a$                                                  & $13.98^{+0.25\,b}_{-0.06}$                                         & $-7.17^{+0.13}_{-0.13}$ & $\approx\!40^b$ & $-2.05 \pm 0.09^c$ & $-2.05 \pm 0.09^c$ & Zn\\
0013$-$004 & $2.09$       & $17.9$   & $1.973$       & $20.83 \pm 0.05^d$                                                  & $18.90^{+1.10\,d}_{-1.10}$                                         & $-1.64^{+1.10}_{-1.10}$ & $\approx\!30^c$ & $-1.75 \pm 0.05^d$ & $-0.93 \pm 0.06^d$ & Zn\\
0347$-$383 & $3.22$       & $17.3$   & $3.025$       & $20.56 \pm 0.05$                                                    & $14.55^{+0.09}_{-0.09}$                                            & $-5.71^{+0.10}_{-0.10}$ & $\approx\!30$   & $-1.72 \pm 0.06$   & $-0.98 \pm 0.09$   & Zn\\
0405$-$443 & $3.02$       & $17.6$   & $2.595$       & $20.90 \pm 0.10$                                                    & $18.16^{+0.21}_{-0.06}$                                            & $-2.44^{+0.23}_{-0.12}$ & $\approx\!30$   & $-1.33 \pm 0.11$   & $-1.02 \pm 0.12$   & Zn\\
0528$-$250 & $2.82$       & $17.2$   & $2.811$       & $21.10 \pm 0.10$                                                    & $18.22^{+0.23}_{-0.16}$                                            & $-2.58^{+0.25}_{-0.19}$ & $[\ga\!40?]^1$  & $-1.26 \pm 0.10$   & $-0.75 \pm 0.10$   & Zn\\
0551$-$366 & $2.32$       & $17.6$   & $1.962$       & $20.50 \pm 0.08^e$                                                  & $17.42^{+0.63\,e}_{-0.90}$                                         & $-2.78^{+0.64}_{-0.90}$ & $\approx\!10^e$ & $-0.96 \pm 0.09^e$ & $-0.13 \pm 0.09^e$ & Zn\\
1232$+$082 & $2.94$       & $18.5$   & $2.338$       & $20.90 \pm 0.10^f$                                                  & $16.78^{+0.10\,f}_{-0.10}$                                         & $-3.82^{+0.20}_{-0.20}$ & $\approx\!10^f$ & $-1.73 \pm 0.13^f$ & $-1.21 \pm 0.15^f$ & Si\\
1444$+$014 & $2.21$       & $18.5$   & $2.087$       & $20.07 \pm 0.07$                                                    & $18.30^{+0.37}_{-0.37}$                                            & $-1.48^{+0.38}_{-0.38}$ & $\approx\!20$   & $-1.58 \pm 0.09$   & $-0.60 \pm 0.15$   & Zn\\\hline
\end{tabular}
Notes: $^1$A value for Q0528$-$250 is not available in
\citetalias{LedouxC_03a} but reasonably high S/N is expected from the QSO
magnitude, exposure time and $z_{\rm abs}$.\\
References: $^a$\citet{LuL_96b}; $^b$\citet{LevshakovS_00c};
$^c$\citet{MolaroP_00a}; $^d$\citet{PetitjeanP_02a};
$^e$\citet{LedouxC_02a}; $^f$\citet{SrianandR_00a};
\end{minipage}
\end{table*}

\subsection{The data}

Table 1 summarizes the properties of the 8 H$_2$-bearing DLAs studied here.
H$_2$ is detected in DLAs via the Lyman and Werner--band UV lines and H{\sc
\,i} is detected via the Lyman series. Observed transitions fall in the
visible region for $z_{\rm abs}$ larger than $\approx\!1.9$.  Column
densities are measured from Voigt profile fits to absorption lines from
each species.  A compilation of results from H$_2$ searches in DLAs is
given in Table 8 of \citetalias{LedouxC_03a}, for which 7 DLAs have
confirmed H$_2$ detections and metallicity measurements: 0013$-$004
($z=1.973$), 0347$-$383, 0405$-$443 ($z=2.595$), 0528$-$250, 0551$-$366,
1232$+$082 and 1444$+$014. The $z=1.973$ DLA towards 0013$-$004 has been
studied in detail by \citet{PetitjeanP_02a}: it comprises several absorbing
components, of which 2 are dominant (their components $c$ and $d$).  These
2 components are distinguishable in H$_2$ but not in the saturated H{\sc
\,i} Lyman series.  $N($H{\sc \,i}$)$ has thus been measured at the mean
redshift of these 2 components and we use the mean $N$(H$_2$) in computing
$f$ for this DLA. The error we use for $f$ is derived from the range in
$N$(H$_2$) given in \citetalias{LedouxC_03a}. We use the value and error
for $N$(H$_2$) in the DLA towards 1232$+$082 from \citet{SrianandR_00a}.

We include one further DLA, 0000$-$263. The H$_2$ detection is regarded as
only tentative in \citetalias{LedouxC_03a} but has been carefully
scrutinised in \citet{LevshakovS_00c,LevshakovS_01b} and relies on two
H$_2$ absorption features, the L(4-0)R1 and W(2-0)Q(1) lines, the former
appearing relatively free from Lyman-$\alpha$ forest blending. Since this
is the highest redshift H$_2$-bearing DLA, the [M/H]--$z_{\rm abs}$ and
$f$--$z_{\rm abs}$ correlations derived below rely on this point relatively
heavily. Further confirmation or refutation of this H$_2$ detection is
therefore desirable. A tentative H$_2$ detection at $z=2.374$ towards
0841$+$129 has also been reported \citep{PetitjeanP_00a}, though
confirmation requires future data and/or analyses.

Here we omit the recent detection towards 0515$-$4414 \citep{ReimersD_03a}
since the local ionizing UV background may be relatively intense. See
\citet*{MurphyM_04b} for an analysis including this system.

All metallicities are based on elements where depletion onto grains is
expected to be small: 7 measurements are derived from [Zn/H], while the
metallicity for 1232$+$082 is derived from [Si/H]. In most DLAs Si is
observed to be slightly dust-depleted with respect to Zn so one might
expect our conclusions below to be slightly strengthened if [Zn/H] could be
used as a metallicity indicator for 1232$+$082.

Fig.~1 illustrates the distribution of [M/H] and $f$ for the 8 DLAs with
detected H$_2$.  The metallicity appears to change by $\sim\!1.5{\rm
\,dex}$ over the range $1.9 < z_{\rm abs} < 3.4$, while the molecular
hydrogen fraction appears to change by $\sim\!6{\rm \,dex}$. The scatter
about the straight line fits is clearly larger than the statistical
(1\,$\sigma$) errors in both panels.  The validity of any simple parametric
fit is thus dubious.  Straight-line fits were nevertheless derived by
quadrature addition of the same constant to each statistical error such
that the normalised $\chi^2 = 1$. Least-squares fits give similar slopes
and $y$-intercepts but our procedure results in reasonable error estimates
for these quantities.

\begin{figure}
\includegraphics[width=\columnwidth]{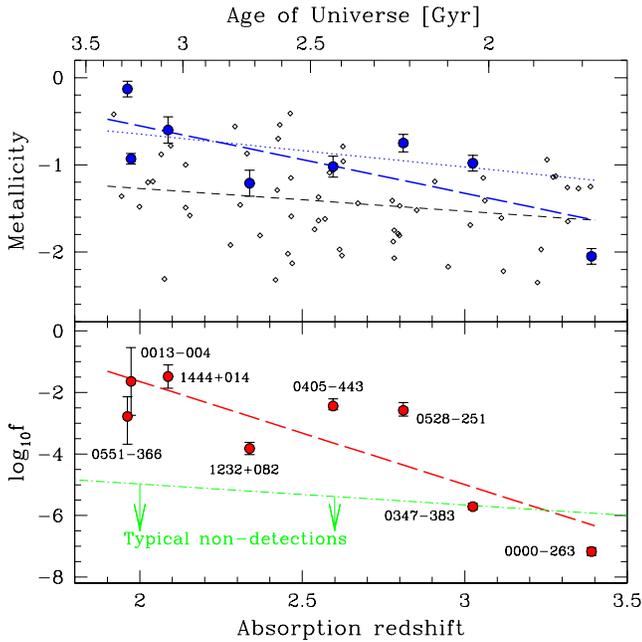}
\caption{{\it Upper panel:} Metallicities for the 8 DLAs which exhibit
H$_2$ absorption (shaded circles), derived from the abundance of Zn (Si for
1232$+$082). The relevant data are summarized in Table 1. The long-dashed
line is a least-squares fit to the 8 points with expanded errors (see
text). The dotted line is a similar fit excluding 0000$-$263. Small hollow
diamonds illustrate metallicity measurements for the 60 DLAs between $1.9 <
z_{\rm abs} < 3.4$ from \citet{ProchaskaJ_03b}. The short-dashed line is
fit to the hollow diamonds, also with expanded errors. {\it Lower panel:}
Logarithmic molecular hydrogen fraction. The dot-dash line represents the
typical non-detection levels found by \citet{LedouxC_03a}. The age of the
universe is calculated using $\Omega_{\rm m} = 0.30, \Omega_{\Lambda}=0.70,
H_0 = 70\,{\rm km\,s}^{-1}\,{\rm Mpc}^{-1}$.}
\end{figure}

\subsection{Metallicity vs. Redshift}

With this procedure, a fit of the form ${\rm [M/H]} = A_8 + {\rm B}_8z_{\rm
abs}$ to the 8 [M/H] points in Fig.\,1 gives ${\rm A}_8=1.0 \pm 0.74$ and
${\rm B}_8 = -0.77 \pm 0.29$. Using Kendall's $\tau$ as a non-parametric
correlation estimate gives $\tau = -0.50$ and $P(\tau)=0.08$. That is,
[M/H] and $z_{\rm abs}$ are anti-correlated with 92\% confidence. Compare
these statistics with those derived using the 60 DLAs from
\citet{ProchaskaJ_03b} in the redshift range $1.9 < z_{\rm abs} < 3.4$
which have Zn, O, S or Si metallicities: ${\rm A}=-0.75 \pm 0.41$, ${\rm B}
= -0.26 \pm 0.15$, $\tau = -0.13$ and $P(\tau)=0.14$. The slope, $B$, also
compares well with that derived by \citeauthor{ProchaskaJ_03b} for the
overall 125 DLA sample, i.e.~$-0.28 \pm 0.05$.

Visually, the [M/H]--$z_{\rm abs}$ distribution suggests two further
results: (i) As discussed in \citet{LedouxC_02a,LedouxC_03a}, the
H$_2$-bearing DLAs have systematically higher [M/H] than the general DLA
population. A Kolmogorov-Smirnov (KS) test reveals the two [M/H]
distributions to have only 2\% probability of being drawn from the same
parent distribution. Removing the lowest and highest redshift H$_2$-bearing
DLAs, which are also the extreme [M/H] points, would give good agreement
between the slope for the remaining 6 points and the general DLA
population, but the metallicity agreement would be somewhat worse (KS
probability reduces to 0.8\%). (ii) The [M/H]--$z_{\rm abs}$ correlation
appears to be tighter for the H$_2$-bearing DLAs compared to the general
DLA population.  To test this we subtract the linear fit to the data in
each case and use an F-test to compare the variances of the residuals.  The
variance for the 60 general DLAs is 1.9 times that for the H$_2$-bearing
DLAs, although the probability of this happening by chance is 39\% given
the sample sizes.

In summary, the H$_2$-bearing DLAs display a faster increase in metallicity
with cosmic time compared to the general DLA population. Furthermore, this
increase is more well-defined in the H$_2$-bearing sample and these DLAs
seem to be chemically distinct from the general DLA population.

\begin{figure}
\includegraphics[width=\columnwidth]{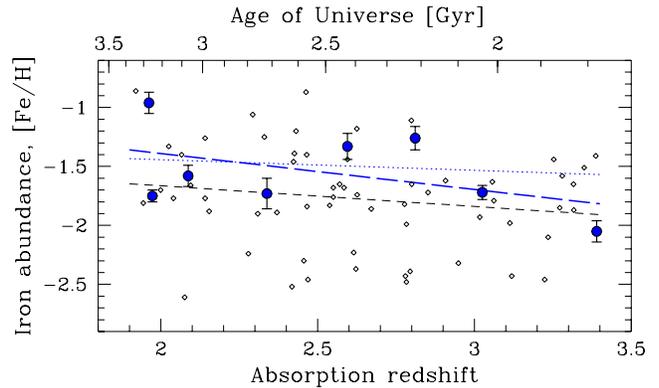}
\caption{As in the upper panel of Fig.~1 but for the distribution of relative
  iron abundance, [Fe/H].}
\end{figure}

These latter two results are also supported by the distribution of relative
iron abundances, [Fe/H], as shown in Fig.~2. Using [Fe/H] for the same 60
general DLAs from \citet{ProchaskaJ_03b} as before, the KS probability for
the two samples being drawn from the same parent distribution is only
0.01\%. The F-test reveals that the variance of the general DLA population
is 2.1 times that of the H$_2$-bearing sample but, again, this has
reasonably low statistical significance since we would expect a larger
variance ratio 29\% of the time by chance alone. Therefore, the
H$_2$-bearing DLAs do seem to form a chemically distinct sub-class of the
DLA population which may encompass a narrower range of physical conditions
and environments. We note in passing that the slope of the [Fe/H]--$z_{\rm
abs}$ relation for the H$_2$-bearing DLAs, in contrast to the
[M/H]--$z_{\rm abs}$ relation, is not significantly steeper than for the
general DLA population. With the same fitting procedure above we find
$A_{\rm Fe} = -0.78 \pm 0.61$ and the slope $B_{\rm Fe} = -0.30 \pm 0.24$
with Kendall's $\tau = -0.21$ and $P(\tau) = 0.46$ for the H$_2$-bearing
DLAs. The general DLA population gives ${\rm A}=-1.3 \pm 0.36$, ${\rm B} =
-0.17 \pm 0.14$, $\tau = -0.10$ and $P(\tau)=0.24$. These shallower slopes
are expected if dust depletion increases in importance with decreasing
redshift, as implied by the $f$--$z_{\rm abs}$ relation discussed below
(Section 3.3). We explicitly treat the evolution of the dust depletion
factor in \citet{MurphyM_04b}.

Finally, it was earlier mentioned that the above conclusions rely
relatively heavily on the highest redshift H$_2$ detection towards
0000$-$263. If we remove this point, the straight-line fits to the
[M/H]--$z_{\rm abs}$ relation give $A_7 = 0.11 \pm 0.81$ and the slope $B_7
= -0.38 \pm 0.33$ with a non-parametric correlation estimate given by
Kendall's $\tau = -0.33$ with $P(\tau) = 0.29$. Thus, the negative
[M/H]--$z_{\rm abs}$ correlation is not as steep, nor as robust when the
0000$-$263 data point is removed. However, note that the evidence
supporting the hypothesis that the H$_2$-bearing DLAs form a chemically
distinct population with a potentially narrower range of physical
conditions is relatively independent of the 0000$-$263 H$_2$
detection. Comparing the 7 remaining H$_2$-bearing DLAs with the 49 general
DLAs between $1.9 < z_{\rm abs} < 3.1$ from \cite{ProchaskaJ_03b} gives a
KS probability of $0.8\%$ and the F-test gives a ratio of variances of 2.6
which occurs 23\% of the time by chance alone.

\subsection{Molecular fraction vs. Redshift} 

Turning to the lower panel of Fig.\,1, we proceed as above and derive a
straight-line fit for the $f$--$z_{\rm abs}$ relation: $A_f = 5.1 \pm 2.4$,
$B_f = -3.4 \pm 0.9$. Kendall's $\tau$ is $-0.50$ with
$P(\tau)=0.08$. Thus, there is reasonable statistical evidence that the
molecular hydrogen fraction increases dramatically with cosmic
time. Previous studies do not explicitly identify this trend.  In
particular, although \citetalias{LedouxC_03a} plot $f$ versus $z_{\rm abs}$
for their sample of DLAs (their figure 16) they do not point out this
apparently rapid evolution of $f$ with $z_{\rm abs}$.

Again, removing 0000$-$263 from consideration weakens the statistical
significance of the $f$--$z_{\rm abs}$ correlation but the slope remains
quite steep: the new fits with 0000$-$263 removed give $A_f = 3.0 \pm 3.0$,
$B = -2.4 \pm 1.2$ with Kendall's $\tau = -0.33$ and $P(\tau) =
0.29$. However, we again stress the apparent reliability of the H$_2$
detection towards 0000$-$263 made by \citet{LevshakovS_00c,LevshakovS_01a}
and point out that H$_2$ systems with high-$z_{\rm abs}$ and high-$f$ (or,
indeed, low-$z_{\rm abs}$ and low-$f$) will need to be detected to show
this correlation to be spurious.

Below we consider possible interpretations of the $f$--$z_{\rm abs}$ trend
but first discuss some systematic effects and biases.

\section{Observational biases and possible systematic effects} 

Could the strikingly steep apparent evolution of $f$ with $z_{\rm abs}$ be
significantly affected by observational selection effects? We consider some
possibilities for both the [M/H]-- and $f$--$z_{\rm abs}$ correlations
below.

Firstly, the H$_2$ sample is inhomogeneous since the quasar spectra do not
all have similar S/N and since the H$_2$ detection methods and criteria
were not uniform. Indeed, the weak H$_2$ lines detected towards 0000$-$263
and 0347$-$383 are at the same level as the many non-detections listed by
\citetalias{LedouxC_03a} (their figure 16). Typically, these fall in the
region illustrated in Fig.~1 by the light dot-dash line. The $f$--$z_{\rm
abs}$ correlation we report is therefore only tentative. However, it seems
surprising that no high-$z_{\rm abs}$/high-$f$ or low-$z_{\rm abs}$/low-$f$
detections of H$_2$ have been made. Secondly, the H$_2$ detection limit
will change with redshift: observed line widths increase as $(1+z)$, making
higher-$z$ lines more detectable, but the Lyman-$\alpha$ forest density
increases roughly as $(1+z)^2$, making higher-$z$ lines harder to detect
due to increased blending. This will be further complicated by higher order
Lyman series forest lines for DLAs at low redshift compared to their
corresponding QSO. Quantification of these competing effects requires
numerical simulations which mimic the (unknown and, likely, non-uniform)
techniques used to detect the 8 H$_2$ systems studied in Fig.\,1.

The [M/H] and $f$ values in Fig.\,1 are not specific to the H$_2$-bearing
components because $N($H{\sc \,i}$)$ is determined from the damped
Lyman-$\alpha$ line profile. The relative abundances of refractory
(e.g.~Fe) and non-refractory (e.g.~Si) elements are generally found to be
uniform across the absorption profiles of most DLAs \citep{ProchaskaJ_03a},
indicating that [M/H] is also likely to be uniform. The notable exceptions
to this rule are the H$_2$-bearing components where large dust depletion
factors are found (e.g.~LPS03 and the [Si/Fe] profile of 0347$-$383 in
\citealt{ProchaskaJ_03a}). These components usually dominate the
non-refractory metal-line profiles and so, although the derived [M/H] and
$f$ measurements will be systematically underestimated, the effect will not
be large. Though the fitted slopes in Fig.\,1 are likely to be reasonably
robust against this effect, a larger sample and more detailed study of the
metal-line and Lyman series profiles is required.

Selection effects related to dust obscuration of quasars may also be
present. In the H$_2$ systems, $f$ is somewhat correlated with dust
depletion factor [Zn/Fe] \citepalias{LedouxC_03a}, providing evidence that
the H$_2$ is formed on dust grains, a view supported by theory
(\citealt{CazauxS_02a,GloverS_03a}, cf.~\citealt{LisztH_02a}). However,
DLAs containing large amounts of dust (i.e.~those with high $f$) could
suppress detection of their background quasars and may therefore be
`missing' from our sample \citep{FallS_89a}.  Since metallicity and dust
depletion are also strongly correlated \citepalias{LedouxC_03a}, this
effect is likely to suppress, rather than create, the correlations in
Fig.\,1. Further to this, \citet{BoisseP_98a} and \citet{PrantzosN_00a}
found an anti-correlation between [Zn/H] and $N($H{\sc \,i}$)$ in DLAs. For
the 8 H$_2$ systems studied here, there exists a mild correlation between
$N($H{\sc \,i}$)$ and $z_{\rm abs}$. If the latter is not simply
fortuitous, it should have resulted in a correlation between $f$ and
$z_{\rm abs}$, opposite to that observed.

The above implies that dust obscuration bias may somewhat {\it suppress}
the observed $f$--$z_{\rm abs}$ correlation. However, several empirical
studies suggest this bias is small. For example, DLA surveys using
radio-selected quasars should not be subject to this bias. Recent work
\citep{EllisonS_01c} indicates that the number of DLAs per unit redshift
interval found in previous optical DLA surveys is not significantly biased,
suggesting the number of `missing' DLAs is at most a factor of 2 below the
radio-selected result. Nevertheless, we note 2 studies which tentatively
suggest strong enhancements in the incidence of absorption towards
radio-loud BL Lac objects \citep{StockeJ_97a} and optically red quasars
\citep{CarilliC_98a}. In the former case, many of the emission redshifts
are not known and so, if intrinsic absorption occurs more frequently in BL
Lacs, the comparison with \citet{EllisonS_01c} is not useful. For the red
quasars of \citet{CarilliC_98a}, only absorption close to the quasar
redshifts was searched for and so, again, the comparison with
\citet{EllisonS_01c} is tentative. Clearly these possible discrepancies
should be resolved observationally in future work.

\section{Interpretations} 

By selecting those DLAs exhibiting H$_2$ absorption, one may focus on
systems with a narrower range of physical conditions than the general DLA
population. Tentative support for this conjecture lies in the tighter
metallicity--$z_{\rm abs}$ anti-correlation observed for the 8 H$_2$ DLAs
studied here. \citet{HouJ_01a} recently presented detailed chemical
evolution models which give a slope for the metallicity--$z_{\rm abs}$
relation of $\approx\!-0.6{\rm \,dex}$. They correct this predicted slope
for various observational biases to achieve the shallower slope observed
for the general DLA population. However, the steep slope observed for 8
H$_2$ systems may be less affected by these biases and may therefore be
more comparable to the uncorrected model slopes. H$_2$-bearing DLAs could
therefore provide an important probe of cosmological chemical evolution.

That there exists such a large range ($\sim\!6{\rm \,dex}$) in the values
of $f$ in Fig.\,1 may not be surprising. For example, \citet{SchayeJ_01b}
describes a photo-ionization model for clouds in local hydrostatic
equilibrium with a representative incident UV background flux, [M/H] and
dust-to-metals ratio. The molecular fraction in this model shows a sudden
increase of $\sim\!4{\rm \,dex}$ for only a small increase in the total
hydrogen density. Therefore, the very steep $f$--$z_{\rm abs}$ correlation
in Fig.\,1 could be achieved with a modest increase in dust content at
lower redshifts, consistent with the [M/H]--$z_{\rm abs}$
correlation. Within the Schaye model, one also expects an anti-correlation
between the intensity of the UV background and $f$. However, the behaviour
of the UV background flux with redshift over the range $z=2$--3 is still a
matter of considerable uncertainty. The strong decrease in $f$ at high
$z_{\rm abs}$ may also be consistent with recent H{\sc \,i} 21-cm
absorption measurements in DLAs \citep{KanekarN_03a}, where a generally
higher spin temperature is found at $z>2.5$. With an increased sample size
and more detailed analyses, the $f$--$z_{\rm abs}$ correlation, if real,
may provide complementary constraints on these problems.

\section*{Acknowledgments}

We enjoyed and benefitted from discussions with Ed Jenkins, Charley
Lineweaver, Mark Whittle and Max Pettini. We thank the anonymous referee
for critical comments which improved the manuscript. SJC gratefully
acknowledges receipt of a UNSW NS Global Fellowship and MTM is grateful to
PPARC for support at the IoA under the observational rolling grant. This
research has made use of the NASA/IPAC Extragalactic Database (NED) which
is operated by the Jet Propulsion Laboratory, California Institute of
Technology, under contract with the National Aeronautics and Space
Administration.


\bsp_small

\label{lastpage}

\end{document}